\documentstyle[twocolumn,aps]{revtex}

\begin{document}

\newcommand{\beq}{\begin{equation}}
\newcommand{\eeq}{\end{equation}}
\newcommand{\bea}{\begin{eqnarray}}
\newcommand{\eea}{\end{eqnarray}}
\newcommand{\ba}{\begin{array}}
\newcommand{\ea}{\end{array}}
\newcommand{\om}{(\omega )}
\newcommand{\bef}{\begin{figure}}
\newcommand{\eef}{\end{figure}}
\newcommand{\leg}[1]{\caption{\protect\rm{\protect\footnotesize{#1}}}}

\newcommand{\ew}[1]{\langle{#1}\rangle}
\newcommand{\be}[1]{\mid\!{#1}\!\mid}
\newcommand{\no}{\nonumber}
\newcommand{\etal}{{\em et~al }}
\newcommand{\geff}{g_{\mbox{\it{\scriptsize{eff}}}}}
\newcommand{\da}[1]{{#1}^\dagger}
\newcommand{\cf}{{\it cf.\/}\ }
\newcommand{\ie}{{\it i.e.\/}\ }
\newcommand{\eg}{{\it e.g.\/}\ }

\title{Contextual objectivity and quantum holism.}
\author{Philippe Grangier}
\address{Laboratoire Charles Fabry de l'Institut d'Optique, %\\
F-91403 Orsay, France}

\maketitle

\begin{abstract}

In our effort to restate a ``realistic" approach to quantum mechanics,
that would fully acknowledge that local realism is untenable, we add
a few more questions and answers to the list presented in quant-ph/0203131. 
We also suggest to replace the very misleading wording ``quantum non-locality"
by ``quantum holism", that conveys a much better intuitive idea of
the physical content of entanglement.

\end{abstract}

\section{Introduction}

In a previous preprint \cite{ph1}, we introduced and discussed a ``physical"
(as opposed to mathematical)  definition of  
a (pure) quantum state, that reads in the following way: 

{\bf The quantum state of a physical system is defined by the values of a 
complete\footnote{The set of quantities is
complete in the sense that the value of any other quantity which satisfies
the same criteria is a function of the set values, see \cite{ph1} for details.}
set of physical quantities, which can be predicted with certainty and measured
repeatedly without perturbing in any way the system}. 

As discussed in detail in  ref. \cite{ph1}, 
this definition is in full agreement with the usual formalism of QM.
It is also implies that some ``objectivity" can be attached to the quantum state,
because the quantum state is defined
from a fully predictable course of events, that is independent of the observer. 
In \cite{ph2} we tried to exploit this definition,
together with some ideas about the system dimensionality,
to propose a new axiomatic approach to QM, that is an attempt
to spell out how the ``quantum reality" is related with the ``macroscopic reality". 
While \cite{ph1} is a straightforward rewriting of usual QM, 
\cite{ph2} is much more tentative and its goal is mostly to stimulate some thinking. 
The FAQ below is a continuation of ref. \cite{ph3}.

\section{More Questions (and answers)}

Q: What about the following  statement, taken from \cite{caf}:
``If you believe that the quantum state is rigidly (or even loosely)
connected to reality, then you will never find a way out of the conundrum of
``unreasonableness" associated with
state-vector collapse at a distance.  I.e., our community will always
be left with a search for the
mechanism that makes it go. Our community will always be left with the
embarrassing questions
to do with its clash with Lorentz invariance. And, maybe most
importantly, we will be left with
the nagging question of why we can't harness this mechanism for more
useful purposes."

A: If we simply admit that there is nothing
like hidden variables, then  Bell's inequalities (BI) can be violated without invoking
anything like an action at a distance, because without the ``$\lambda$'s"
the standard demonstration of these inequalities does not hold. 
Then in a classical view there should be no
correlations either, but this only proves that classical theory does not work.

The apparent ``action at a distance" only results from the effort of attributing a
- non-existent - ``reality" to the state of a sub-part of a system. 
The conclusion taken in the quotation above is
that ``one cannot connect the quantum state to reality". But this is {\bf not}
correct : the correct conclusion is that a subpart of the system has {\bf no}
quantum state, while the overall system (the pair) certainly has one,
because it has a full set of properties that can be predicted with certainty.
To state it again:

- let us admit that there is a ``reality" attached to the pair of particles, but
that there is no ``reality" attached to each particle (here ``reality" simply
means a quantum state according to our definition).

- the ``reality" attached to the pair makes no problem with Lorentz
invariance, because it was created when the two particles interacted,
and it simply follows them if they move very far away. The same
conclusion apply to more fancy schemes like entanglement swapping, that
requires classical communications to effectively prepare the remote
entangled state.

- since there is no ``reality" attached to each particle, Bell's
inequalities do not hold !  All the job is done by the fact that the
individual particles have no quantum state, or no other property
whatsoever that would decide on the result of the measurement. Then
``action at a distance" simply vanishes: if a measurement if performed on
one side, {\bf absolutely nothing} happens on the other side. 

As far as ``action at a distance" is concerned, the situation is just
the same as in the classical example where I mail a red and a white
paper in two envelopes, one to New Jersey and the other
one to Australia : if everybody trust my ``state
preparation" (many witnesses can swear that it is ``real"), the guy in Bell Labs will
know immediately the colour of the other paper when opening  his envelope.
The shocking question is :  but ``what" carries the correlation if the
``papers" have no ``colour", as it is the case in QM ? The answer is : the
correlation is carried by the quantum state of the pair, that is real
and that propagates causally. All the trouble here comes from the
desperate attempt to map this quantum evidence on a classical picture,
that is out of the game. 

Actually, what I am saying is old hat : there is a well known quotation
by Einstein (in approximate form) ``There is no doubt that the physicists
who adhere to QM will react in the following way : they will give up the
requirement of the separate existence of physical reality in different
parts of space; they will be right to say that QM makes no explicit use
of this requirement". Then Einstein wrote that this was 
unacceptable to him, 
and maybe this is still so difficult to accept because 
it was said many time that the price to pay is giving up physical
reality. But this is wrong : in our point of view local realism was an
inheritage from classical physics, that can be abandoned without giving
up neither locality (in the sense of relativistic causality), nor
realism (in the sense of the existence of an objective physical reality). 
We can thus admit now that local realism is dead, but that physical
realism can do quite well without it.  

What next ? QM is a fantastic theory, and it seems relevant to ask:
why is it working so well ? Actually, QM was
invented 75 years ago in a somehow anarchic way, as an attempt to
understand atomic spectra. We may thus speculate the following: QM is
actually the answer to a question that was never clearly formulated. We
have the answer, what about finding the question ? 

%\vskip 2mm

Q: What about nonlocality without entanglement \cite{chb} ? 

A: In \cite{chb} the authors present a set of orthogonal product states of two three-states
particles, that cannot be reliably distinguished by a pair of separate
observers who are given one particle each, and who are
ignorant of which of the states has been presented to them
(the separate observers are allowed any sequence of local operations
and classical communications). This was termed
``quantum nonlocality without entanglement" in \cite{chb},
and ``EPR paradox without entanglement" in \cite{hhh}. This latter paper
states that ``negation of local realism is roughly called non-locality",
and that ``informational local realism" is the concept that is at stake
in this example.  

First we don't fully agree with the statement  
``negation of local realism is roughly called non-locality".
As said above, the negation
of local realism is that it is not possible to define objectively the quantum
state of a subpart of a quantum system; there is no need, and it may even be
misleading, to call this  non-locality. 
We fully acknowledge however that the most straightforward way to 
demonstrate that local realism is contradictory with QM is by using Bell's inequalities,
that use entangled states (on the quantum side) and locality (on the
classical side). But once it is admitted that QM has nothing to do
with local realism, there is no reason to restrict this property to entangled states.
The situation considered in \cite{chb} gives a striking example that
the joint state of a quantum system can be given more ``reality" 
(\ie more predictability and reproducibility) than
the ``states" of its  parts. The state of the joint system can indeed
be determined with certainty by performing a joint measurement
(in the appropriate measurement basis),
while separate measurements done on the parts cannot provide an equivalent
result, whatever is done on both sides. 
%So now the question should be:
%if non-locality is innocent, who framed local realism ? 

%\vskip 2mm

Q: Can contextual objectivity say anything about the quantum state of the universe 
in a cosmological sense ?

A: Quantum cosmology is a subject that is far beyond my area of knowledge. However,
here is an example of statement (taken from \cite{qu}),
that enters in contradiction with contextual objectivity:
``The first important question is this: is the universe a complete quantum system or not ?
By complete we mean that there are no external observers or classical agencies not subject
to the laws of quantum mechanics.
The observation of Bell-type correlations and violations of Bell-type inequalities together 
with a vast amount of empirical evidence for the universal correctness of quantum principles 
strongly suggests that the answer to the first question is yes. This answer excludes from 
further discussion all paradigms and models based on any sort of classical hidden variables. 
We assume henceforth that the entire universe is a vast, self-contained quantum automaton".

In our view, the authors' conclusion is unwarranted. %We consider that
Bell's theorem proves that there are no ``classical hidden variables" to be looked for
``below" or ``inside" quantum mechanics (QM). But it does not prove that QM is an 
``embedding theory", i.e., that ``the universe a complete quantum system" in
the authors' sense. There is another possible conclusion, i.e. to say that the universe has
both a classical structure {\bf and} a quantum sub-structure. But it may have neither a  
classical sub-sub-structure (hidden variables), nor a quantum super-structure (multiverse
or other versions of a fully quantum universe). 

This may be shocking in the sense that we as physicists are strongly tempted to say : ``if QM
is right, then it must be universal". But one has too look carefully at what ``universal" means.
It certainly means that every time one looks at a quantum system, i.e. a system
where quantization plays an important role, then this system appears to be quantum. 
But it certainly does not mean 
that linear superpositions and so forth are an obvious feature of our everyday's environment.
It may be possible to claim that ``we cannot see such superpositions, but even
if they have undergone a decoherence process they
must remain around there, because the overall evolution must be unitary". %\cite{dd}
Our approach is still different, 
since we claim that QM is universal, in the sense that it
provides the only way to relate our classical environment, that is essentially based upon
continuity, and its quantized sub-structure \cite{ph2}. 
{\bf In other terms, QM is basically an interface tool}.

We don't conclude either that a quantum state is ``purely subjective", or ``a state of our mind",
or whatever similar. A (pure) quantum state is a fairly objective status for a quantum system,
in particular because there is no ``ignorance" left (a pure state has zero entropy). 

So there is clearly some conflict between contextual objectivity and the more
theoretical view that ``QM is either everything or nothing". 
I am not able to spell out whether or not the present
approach may have some implications for cosmology or quantum gravity, but hopefully it may
interest people competent in these fields.

%\vskip 2mm

Q: What about ``quantum holism" ?

A: For explaining to the layman what is quantum entanglement, you may say either :

(i) when two remote particles are in an entangled state, there is some kind
of mysterious influence or instantaneous action-at-a distance between them, as it 
is proven by the observed violation of Bell's inequalities. Though this 
action-at-a-distance cannot be used for superluminal signalling,
it is a clear evidence for quantum non-locality.

or :

(ii) quantum entanglement is the fact that in quantum mechanics it is
possible to attribute definite predictable properties to an ensemble of
several particles, while it is impossible to do so for each of the particles
taken separately. This impossibility contradicts the classical assumption 
that the state of a given composite system can be defined from the state of its parts.

It should be clear to the reader that in our approach, (i) is misleading and 
even wrong, while (ii) is perfectly fine, and moreover is in perfect agreement
with the mathematical content of the quantum formalism. But for an unknown 
reason, (i) usually generates a lot 
of excitement, while (ii) will generate no excitement at all - simply a glance
of deep misunderstanding (you may try it, I did). 

Therefore I propose to use the words ``quantum holism" to convey the idea that 
(ii) {\bf is} strange, 
interesting and very far from classical physics. This wording expresses the ability 
of the quantum formalism to attribute ``more reality"
to a composite system than to its parts. Then (ii) can be rewritten as~:

(iii) quantum entanglement is the fact that quantum mechanics 
can attribute fully predictable properties to an ensemble of
several particles, in such a way that knowing these properties
is {\bf contradictory} with knowing predictable properties for each of the particles. 
This ``quantum holism" strongly contradicts the classical assumption 
that the state of a given composite system can always be reconstructed
from knowing the states of its parts.

From the experience of anyone who has been reviewer and editor of
scientific papers, or simply reader of broad-audience essays about quantum mechanics, 
the number of irrelevant or even wrong discussions that have been triggered by the
wording ``quantum non-locality" is simply astonishing. So it may be worthwhile
to give a try to ``quantum holism", that has essentially the same 
content for those who know and use quantum mechanics, but 
may convey quite different ideas. And it does say what we wish to say~:
quantum mechanics has {\bf no} conflict with causality or relativity,
but it gives an holistic description of the state of a physical system. 
Quantum holism is the true ``quantum scandal", that is at the heart of entanglement,
EPR paradox, violation of Bell's inequalities, and decoherence. 
After 75 years of quantum mechanics it probably deserves some recognition. 

\vskip 1.5cm

Q: Are not you unfair to Chris Fuchs' position ?

A: The sentence from \cite{caf} used as the first question of the present paper
may be quoted in an inappropriate way indeed~: 
its initial purpose was to illustrate the state
of affair before Bell's theorem, and not to express 
Chris Fuchs' personal views on quantum mechanics (thanks to David Mermin
for pointing that out). 

Nevertheless, if I remember well a discussion about Chris Fuchs' position \cite{SvE}, 
the conclusion was that a quantum state is not objective, 
but that it is not less  objective than a classical state. 
If this is the case, the debate vanishes~: 
my point is not to do philosophy about the ultimate nature of reality, 
neither to be unfair to Chris Fuchs, but to criticize and if possible to 
dismiss the often-heard statement that classical physics is the domain 
of realism and objectivity, while quantum physics would be the domain 
of fogginess and subjectivity \cite{ew}. 

I do that by claiming that the EPR criterion of reality is (almost) valid in QM, 
with the caveat that it cannot be applied to all physical quantities simultaneously, 
but only to a 
subset of them. Nevertheless, this is strong enough to attribute (contextual) 
objectivity to a quantum state. 
The characteristic feature of an entangled state is 
that ``reality" can be attributed only to joint physical properties, 
and not to individual ones. This is why I think that the usual wording 
of ``quantum non-locality" that is 
attributed to entanglement is very misleading, 
and should be replaced by ``quantum holism".

Actually, Bell's theorem proves that if a ``classical" theory 
({\it i.e.}, a theory where all particles have definite physical properties, 
that may be unknown to us) wants to reproduce 
QM and be in agreement with the experiments, then it must be non-local 
({\it i.e.}, it must include ``instantaneous action at a distance"). 
But does that imply that QM itself requires anything
happening instantaneously at a distance ? My answer is definitely {\bf no}. 

It is certainly generally admitted that 
QM does not allow super-luminal communications. 
But then it is often said that QM include 
mysterious ``non-local influences", often by invoking  
the ``instantaneous reduction of the wave packet". This is nonsense. 
Bell's inequalities are violated because the {\bf first}
Bell's hypothesis  fails, {\it i.e.} that, in contradistinction
with classical physics, QM is able to describe global properties 
that are {\bf not} carried by properties of individual particles : 
this is what I call quantum holism. 

But is not quantum holism simply the same thing as ``quantum non-locality" ? 
Yes indeed, if ``quantum non-locality" is understood correctly, 
recognizing that it has nothing to
do with anything ``instantaneous at a distance". But if the meaning of
``quantum non-locality" is unclear to you,
quantum holism has the huge advantage that it does not suggest any kind of 
``instantaneous transmission", but explicitly refers to the existence of 
global properties that are not contained in the properties of the subparts. 
This is in my opinion the 
correct physical view that is underlying entanglement, 
EPR paradox, Bell's inequalities and even decoherence.


\begin{references}
\bibitem{ew} An even worse statement is that the only possible way to make QM
understandable would be to make it ``classical-looking", either by re-introducing
some kind of hidden variables, or by showing that Bell's theorem is ``wrong".
\end{references}
\end{document}